\documentclass[12pt,letterpaper]{article}
\pdfoutput=1
\usepackage{amsmath,amsfonts,amssymb,mathtools}
\usepackage[nosort]{cite}
\usepackage{graphicx}
\usepackage{jheppub}
\usepackage[T1]{fontenc} 
\usepackage{epsfig}
\usepackage{psfrag}	
\usepackage{tikz}
\usepackage{slashed}
\usepackage{tensor}
\usepackage{ulem}
\usepackage[font=small,labelfont=bf,width=1\textwidth]{caption}
\usepackage{comment}
\usepackage[all]{xy}
\usepackage{multirow}
\usepackage{float}
\usepackage{cite}
\usepackage{color}

\usepackage{tocloft} 
\usepackage{ulem}
\usepackage{soul,xcolor}
\setstcolor{red}

\numberwithin{equation}{section}

\def\ZK#1{{\color [rgb]{0,0.6,0} [ZK: #1]}}

\DeclareMathOperator{\Tr}{Tr}

\newcommand{\bea}{\begin{eqnarray}}
\newcommand{\eea}{\end{eqnarray}}
\newcommand{\beq}{\begin{equation}}
\newcommand{\eeq}{\end{equation}}
\newcommand{\bal}{\begin{equation}\begin{aligned}}
\newcommand{\eal}{\end{aligned} \end{equation}}

\newcommand{\cF}{{\mathcal F}}

\newcommand{\cN}{{\mathcal N}}

\title{The planar limit of integrated 4-point functions}

\author[a]{Bartomeu Fiol}
\author[b]{and Ziwen Kong}

\affiliation[a]{Departament de F{\'\i}sica Qu\`antica i Astrof\'isica i \\Institut de Ci{\`e}ncies del Cosmos, 
Universitat de Barcelona,
Mart{\'\i}\ i Franqu{\`e}s 1, 08028 Barcelona, Catalonia, Spain}
\affiliation[b]{Department of Mathematics, King's College London,
\\
The Strand, WC2R 2LS London, United-Kingdom}

\emailAdd{bfiol@ub.edu}
\emailAdd{ziwen.kong@kcl.ac.uk}

\abstract{We compute the planar limit, as all-order power series in the 't Hooft coupling, of various integrated 4-point functions of chiral primary operators of ${\cal N}=4$ SU(N) super Yang-Mills, and of moment map operators of ${\cal N}=2$ SU(N) SQCD. We do so by computing the planar free energy on $S^4$ of the respective massive deformations of these theories,  and then taking advantage of the exact relation between these free energies and the integrated 4-point functions.}

\begin{document}
\maketitle

\section{Introduction}
N-point functions of local operators are among the basic objects of interest in any given quantum field theory. For generic quantum field theories, it is exceedingly difficult to compute them exactly. In theories with extended symmetry - {\it e.g.} conformal and/or supersymmetry - and for selected operators, there are additional tools that greatly enhance our ability to determine these n-point functions. This paper is devoted to a couple of such examples; let's introduce them in turns.

Among four-dimensional quantum field theories, ${\cal N}=4$ super Yang-Mills (SYM) theories with arbitrary gauge group G play a distinguished role, having maximal supersymmetry (without gravity) and being conformal for any gauge group. Moreover, for classical Lie groups G, the AdS/CFT duality allows to compute certain n-point functions of this theory in regimes hard to access with ordinary field theoretic techniques.

The ${\cal N}=4$ superconformal algebra contains a set of distinguished operators grouped in 1/2 BPS short multiplets, whose superconformal primaries we will denote by $S_p$ and have conformal dimensions $\Delta=p$ \cite{Dolan:2002zh}. The stress-energy tensor belongs to the same multiplet as $S_2$, making this multiplet particularly relevant physically. The 2- and 3- point functions of these operators are not renormalized {\it i.e.} they are independent of the coupling, and can be computed in the free theory \cite{Baggio:2012rr}. Thus, the first n-point functions of these operators with non-trivial dependence on the marginal coupling are 4-point functions (see \cite{Heslop:2022qgf} for a recent review of these 4-point functions). For instance, this includes the stress-energy tensor  4-point function, which is dual to the  $2\rightarrow 2$ graviton scattering in $AdS_5$. A subset of these 4-point functions will be the first focus of attention in this paper. 


The second example that we will consider concerns ${\cal N}=2$ superconformal field theories (SCFTs). Four dimensional superconformal field theories with ${\cal N}=2$ supersymmetry have a plethora of short and semishort multiplets \cite{Dobrev:1985qv, Dolan:2002zh}. Among them, two particular families of short ${\cal N}=2$ multiplets stand out: ${\cal E}_r$ (or Coulomb type) and ${\hat {\mathfrak{B}}}_R$ (or Higgs type) multiplets\footnote{Arguably, an even more important ${\cal N}=2$ short multiplet is the one containing the ${\cal N}=2$ stress-energy tensor. In this work we will have nothing to say about n-point functions of operators belonging to this multiplet.}. The superconformal primaries of ${\cal E}_r$ are annihilated by supercharges of the same chirality, and we will refer to them as chiral primary operators. On the other hand, superconformal primaries of ${\hat {\mathfrak{B}}}_R$ multiplets are annihilated by supercharges of the opposite chirality. In particular,  conserved currents of global symmetries belong to the simplest non-trivial such multiplet,  the moment map multiplet ${\hat {\mathfrak{B}}}_1$.

Correlation functions of ${\cal N}=2$ chiral primary operators have been studied from many different perspectives, particularly in the case of extremal correlators, which involve $n-1$ chiral primary operators and just one anti-chiral primary one \cite{Papadodimas:2009eu}. As for ${\cal N}=2$ Higgs type operators, their 3- point functions are non-renormalized for specific kinematics \cite{Beem:2013sza} and their 4-point functions have been discussed using various techniques \cite{Eden:2000qp,  Nirschl:2004pa, Dolan:2004mu, Beem:2014zpa}. These 4-point functions will be the second main focus of attention in this paper.

A very powerful technique to compute certain observables in ${\cal N}=2$ SYM theories is supersymmetric localization \cite{Pestun:2007rz}, so it is natural to inquire whether it can be applied to the computation of n-point functions of the operators listed above. It turns out that this can be accomplished by considering deformations of the original theory that preserve ${\cal N}=2$ Poincar\'e supersymmetry, and then applying supersymmetric localization to the resulting deformed theory. There are two types of such terms \cite{Argyres:2015ffa,Cordova:2016xhm}: deformations associated to Coulomb branch chiral primaries, and mass deformations associated to the moment map multiplet.

As a first realization of this idea, the authors of \cite{Gerchkovitz:2016gxx} considered deformations of ${\cal N}=2$ Lagrangian SCFTs by top components of chiral multiplets. They showed that extremal correlators of the corresponding chiral primaries can be obtained by taking derivatives of the free energy of the deformed theory, which in principle can be computed exactly by supersymmetric localization.
Remarkably, while the deformations involve integrated correlators of top components of the chiral multiplets, it can be shown that they are equivalent to {\it unintegrated} correlators of the chiral primaries \cite{Gomis:2014woa, Gerchkovitz:2016gxx}.


The second type of deformations are mass deformations. It has been recently realized that these deformations allow to relate integrated 4-point functions of the original SCFT with the free energy of the massive theory on $S^4$. Specifically, certain integrated 4-point functions of ${\cal N}=4$ SYM chiral primaries are related to the free energy on $S^4$ of a massive deformation of ${\cal N}=4$ SYM, ${\cal N}=2^*$, further deformed by chiral multiplets \cite{Binder:2019jwn, Chester:2020dja}. More generally, integrated 4-point functions of moment map operators of ${\cal N}=2$ SCFTs are related to the free energy of the massive theory \cite{Chester:2022sqb}. 

The goal of the present work is to compute the planar free energy of some massive deformations of ${\cal N}=2,4$ SCFTs, and thus obtain the corresponding integrated 4-point functions in the planar limit. All the quantities considered in this work admit a double expansion, as a perturbative series in the 't Hooft coupling $\lambda$ around $\lambda=0$ and in the transcendentality of the coefficients. As we will argue in the main body of this work, for some quantities the relevant matrix models are particular cases of the family solved in \cite{Fiol:2020bhf, Fiol:2020ojn, Fiol:2021icm} in the planar limit, while for others we have to extend the methods developed in \cite{Fiol:2020bhf, Fiol:2021icm}.  

The structure of the paper is as follows. Section 2 is devoted to set the stage for the rest of the paper: we describe the operators that we are going to consider; we recall the relations between their integrated 4-point functions and the free energy of the corresponding massive theories, and finally we explain how to compute these free energies in the planar limit.

In section 3 we look at the first specific case, that of 1/2 BPS chiral primary operators of ${\cal N}=4$ SU(N) SYM. The relevant fourth derivatives of the planar free energy that we find are

\bal
-\frac{\partial_{\tau_p} \partial_{\bar \tau_p} \partial^2_m {\cal F}}{\partial_{\tau_p} \partial_{\bar \tau_p} {\cal F} } \bigg|_{m=0}=
4p\sum_{n=1}^\infty \zeta_{2n+1} (-\tilde \lambda)^n (2n+1){2n\choose n}
\left[(-1)^p{2n \choose n+p}+{2n \choose n+1}\right]
\label{four1intro}
\eal

\bal
&-\partial^4_m {\cal F}\vert_{m=0}= 2\sum_{n=1}^\infty \zeta_{2n+1} (-\tilde \lambda)^{n-1} (2n+1) {2n \choose n} {2n \choose n+1} + \\
& 48 \sum_{m,n=1}^\infty \zeta_{2m+1}\zeta_{2n+1}(-\tilde \lambda)^{m+n}\frac{(2m+1)!}{m!m!}\frac{(2n+1)!}{n!n!} \sum_{i=1}^m {m\choose i}{m \choose i-1} \sum_{j=1}^n {n\choose j}{n \choose j-1} \frac{1}{i+j}
\label{four2intro}
\eal
with $\tilde \lambda =\lambda/16\pi^2$, and $\tau_p, \bar \tau_p$ the parameters for chiral and antichiral deformations. It is readily checked that (\ref{four1intro}) is the perturbative expansion around $\lambda=0$ of the result found in \cite{Binder:2019jwn}. It is worth noting that (\ref{four1intro}) is rather similar  - but not identical - to the terms with maximal transcendentality of ${\cal N}=2$ SU(N) SQCD planar 2-point functions of chiral operators \cite{Fiol:2021icm, Fiol:2022vvv}. As for (\ref{four2intro}), the authors of \cite{Chester:2020dja} derived an expression for $\partial_m^4 {\cal F}\vert_{m=0}$ in terms of single and double integrals (see eq. (A.24) in \cite{Chester:2020dja}). We haven't proved that its perturbative series is given by (\ref{four2intro}), but to the order that we have checked, they do agree.

In section 4, we turn to the case of 4-point functions of moment map operators in ${\cal N}=2$ SU(N) SQCD, {\it i.e.} SU(N) with $N_F=2N$ massless hypermultiplets in the fundamental representation. Already prior to the massive deformation, the planar free energy of this theory is vastly more complicated than that of ${\cal N}=4$ SU(N) SYM \cite{Fiol:2020bhf}. This planar free energy can be organized as a double expansion in $\lambda$ and in the transcendentality of the coefficients for a fixed power of $\lambda$. These coefficients can be expressed in a purely combinatorial fashion, in terms  of a sum over tree graphs \cite{Fiol:2020bhf}.

We argue that there are two independent massive deformations to consider; however, we find it convenient to write the planar free energy in terms of three contributions
\bal
{\cal F}=N \alpha_1 (\lambda) \, \sum_{u=1}^N M_u^4 +\alpha_2 (\lambda) \sum_{u\neq v} M_u^2 M_v^2 + N^2\alpha_3(\lambda) \, \left( \sum_{u=1}^N M_u^4-\frac{1}{N-1} \sum_{u\neq v} M_u^2 M_v^2\right)
\label{intro3alphas}
\eal
In the large N limit, the term with $\alpha_3(\lambda)$ is dominating. In the main text we will argue that the subleading corrections to $N^2\alpha_3(\lambda)$ are of order $N^0$, so it is meaningful to keep $\alpha_1(\lambda),\alpha_2(\lambda)$ as the first non-planar corrections. The methods of \cite{Fiol:2020bhf, Fiol:2021icm} provide closed expressions for $\alpha_1 (\lambda), \alpha_2(\lambda)$ to all orders in $\alpha$ and in transcendentality. As for $\alpha_3 (\lambda)$, we present an algorithm to compute all the terms of a given value of transcendentality, and evaluate the leading and subleading terms, to all orders in $\lambda$. Currently we lack a neat combinatorial interpretation of the resulting perturbative series for $\alpha_3(\lambda)$.

Finally, an appendix contains some technical details of an expansion performed in section 3.

There are a number of possible generalizations of the present work, let's mention a few. First, it would be nice to find a complete characterization of the double perturbative series for $\alpha_3(\lambda)$ in (\ref{intro3alphas}). This will require going beyond the methods developed in \cite{Fiol:2020bhf, Fiol:2021icm}, and it might find applications in the evaluation of other ${\cal N}=2$ observables in the planar limit. A possible application of the present work is to evaluate 4-point functions of Higgs branch operators of ${\cal N}=2$ SCFTs with known holographic duals. This might pave the way to provide a check of holography in the Higgs branch. Conceptually this ought to be straightforward, as the relations found in \cite{Chester:2022sqb} apply to any ${\cal N}=2$ SCFT.

\section{Integrated 4-point functions and localization}

In this section, we introduce all the ingredients needed in the rest of the paper. First, we are going to review the operators that we will consider, and their realization in terms of ${\cal N}=2$ super Yang-Mills fields. Then, we will briefly review how to relate integrated versions of their 4-point functions to derivatives of the free energy of massive ${\cal N}=2$ theories on $S^4$. Finally, we will recall how supersymmetric localization reduces the evaluation of the free energy to a matrix integral, and how in some cases we can efficiently compute its planar limit.

\subsection{Operators}
To start, recall that the ${\cal N}=2$ superconformal algebra in four dimensions is $su(2,2|2)$ \cite{Dolan:2002zh}. Its maximal bosonic subalgebra is $so(4,2)\times su(2)_R\times u(1)_r$, so its irreducible representations are labelled by the quantum numbers of the superconformal primary: $(\Delta; j_1,j_2; R,r)$. Unitarity bounds translate into inequalities that these quantum numbers must satisfy, and superconformal primaries saturating these inequalities define short multiplets \cite{Dolan:2002zh}. Two such short ${\cal N}=2$ multiplets stand out:

\begin{itemize}
\item{ ${\cal E}_r$, 1/2 BPS Coulomb type multiplets. Assuming that their superconformal primary is a Lorentz scalar, it has quantum numbers $(\Delta,0,0;0,\Delta)$. }
\item{${\hat {\mathfrak{B}}_R}$, 1/2 BPS Higgs type multiplets. Their superconformal primary has quantum numbers $(\Delta,0,0;\Delta/2,0)$. The case $\Delta=2$ is particularly important, since the multiplet contains conserved currents of the global symmetries. This is the moment map multiplet, ${\hat {\mathfrak{B}}_1}$.}
\end{itemize}

In the particular case where the ${\cal N}=2$ SCFT is a ${\cal N}=2$ super Yang-Mills theory, these multiplets are realized as follows. The ${\cal N}=2$ vector multiplet contains a complex scalar field $\phi$, and single trace operators tr $\phi^p$ are superconformal primaries of ${\cal E}_p$ multiplets. The ${\cal N}=2$ hypermultiplet contains a $SU(2)_R$ doublet of complex scalar fields, $Q_{I\,i}^a$, with $I,J=1,2$ a $SU(2)_R$ index, $a=1,\dots,N_c$ and $i=1,\dots, N_f$. The superconformal primaries of the moment map multiplet are bilinears of these scalars.

We will also consider certain short multiplets of ${\cal N}=4$ SYM. In this case, the R-symmetry algebra is $SU(4)_R$, so multiplets are labelled by $(\Delta;j_1,j_2;k_1,k_2,k_3)$ with $[k_1,k_2,k_3]$ labelling an irreducible representation of $su(4)$. A particularly interesting family of short representations has superconformal primary operators in the $[0,p,0]$ representation of $SU(4)_R$ with protected conformal dimension $\Delta=p$. In this work, superconformal primaries of these multiplets are realized as single traces of the six real scalars $X^I$, $I=1,\dots,6$ of the ${\cal N}=4$ SYM theory

\bal
S_p(x) = C_{I_1\dots I_p} \text{ tr } X^{I_1}\dots X^{I_p}
\label{thespops}
\eal
with $C_{I_1\dots I_p}$ symmetric traceless $so(6)$ tensors.
For our purposes, it is convenient to think about ${\cal N}=4$ SYM in ${\cal N}=2$ language. The $SU(4)_R$ symmetry decomposes then into $SU(2)_F\times SU(2)_R\times U(1)_r$. Then, the 1/2 BPS $[0,p,0]$ ${\cal N}=4$ multiplet contains, among others, a chiral and an antichiral ${\cal N}=2$ multiplet with $\Delta=p$, and a ${\hat {\mathfrak{B}}}_{p/2}$ multiplet \cite{Dolan:2002zh}. In particular, the ${\cal N}=4$ $[0,2,0]$ multiplet contains a ${\cal N}=2$ moment map multiplet.

\subsection{4-point functions}
Let's start discussing 4-point functions of $S_p$ operators in (\ref{thespops}). The methods of \cite{Binder:2019jwn} apply to a particular subset of 4-point functions of these operators, namely, to $\langle S_2(x_1)S_2(x_2) S_p(x_3) S_p(x_4)\rangle$. These 4-point functions can be written in terms of a single non-trivial function ${\cal T}_p(U,V)$ of the conformal cross-ratios $U,V$ \cite{Dolan:2001tt} (see \cite{Binder:2019jwn} for a recent discussion), and our goal will be to obtain perturbative series for integrals involving this function. 

Let's recall now the structure of the 4-point function of moment-map operators \cite{Dolan:2001tt, Beem:2014zpa}. Introduce an $SU(2)_R$ polarization $y$, and let $A=1,\dots,$ dim G be an index in the adjoint of the flavor symmetry group G. Then, the 4-point function takes the form
\bal
\langle \phi^A (x_1,y_1) \phi^B (x_2,y_2) \phi^C(x_3,y_3) \phi^D(x_4,y_4)\rangle =
\frac{\langle y_1,y_2\rangle^2 \langle y_3,y_4\rangle^2}{(x_1-x_2)^4 (x_3-x_4)^4} G^{ABCD}(U,V;w)
\eal
with
\bal
w=\frac{\langle y_1, y_2 \rangle \langle y_3,y_4\rangle}{\langle y_1,y_3 \rangle \langle y_2,y_4 \rangle }
\eal
Moreover, $G^{ABCD}$ can be decomposed in flavor projectors $P_r^{ABCD}$ where $r$ runs over the irreducible representations that appear in the tensor product of $adj\, G$ with itself
\bal 
G^{ABCD}(U,V;w)=\sum_{r\in \text{Adj} \otimes \text{Adj}} G_r^{ABCD} (U,V;w) P_r^{ABCD}
\eal
Finally, all non-trivial information contained in $G_r^{ABCD}(U,V;w)$ is encoded in an R-symmetry singlet ${\cal G}_r^{\text{int}}(U,V)$ \cite{Chester:2022sqb}. Again, our goal is to derive the planar limit of integrals involving ${\cal G}_r^{\text{int}}(U,V)$. 

\subsection{Relation with free energy}
Let's now briefly review the relation between 4-point functions of operators introduced above, and the free energy of massive deformations of the corresponding SCFT defined on $S^4$. Let's start with Coulomb branch deformations. It will be important to differentiate conformal primaries on $S^4$, that we denote by $\Omega_p$, and those on $\mathbb{R}^4$, that we denote by ${\cal O}_p$. This kind of deformation involves an integral over the top component of the $\Delta=p$ Coulomb branch operator \cite{Gerchkovitz:2016gxx}
\bal
S_p=\tau_p \int d^4x\, {\cal Q}^4\Omega_p
\eal
Remarkably this turns out to be equivalent to inserting the superconformal primary $\Omega_p$ on the North pole of $S^4$, so we can write it as \cite{Gomis:2014woa, Gerchkovitz:2016gxx}
\bal 
S_p=\tau_p \Omega_p (N)
\eal
and similarly, the insertion of an antichiral multiplet is equivalent to the insertion of its superconformal primary at the South pole of $S^4$.

The second type of deformation involves an integral over the moment map multiplet. In terms of ${\cal N}=2$ SYM fields, it adds a mass term for ${\cal N}=2$ hypermultiplets. If the flavor group is G, and $A=1,\dots,dim\, G$ an index in the adjoint representation of $G$
\bal
S_m =\int d^4x \, \sqrt{g} \left( m_A \left[\frac{i}{r}J^A+K^A\right]+m^2 L\right)
\eal
where $J$ is a $SU(2)_R$ invariant, bilinear in the superconformal primaries of ${\hat {\mathfrak{B}_1}}$. It has dimension 2, and it appears divided by the radius $r$ of $S^4$; this term does not appear in $\mathbb{R}^4$. $K^A$ is a scalar of dimension 3, bilinear in the fermions of the ${\cal N}=2$ hypermultiplet. The $m^2$ corresponds to the $\mathbb{R}^4$ mass term for the bosons of the ${\cal N}=2$ hypermultiplet, and won't play any role in what follows. Contrary to what happened in the previous case, the integrated deformation doesn't simplify to an unintegrated insertion. 

In \cite{Binder:2019jwn, Chester:2020dja} it was argued that taking derivatives of the free energy of the deformed theories with respect to the deformation parameters, one obtains exact relations involving n-point functions of the deformation integrals above. Using conformal symmetry, it is possible \cite{Binder:2019jwn} to rewrite the relation in terms of just an integral over ${\cal T}_p(U,V)$

\bal
l_p=\frac{\partial^2_m \partial_{\tau_p} \partial_{\bar \tau_p} \log \, Z}
{\partial_{\tau_p} \partial_{\bar \tau_p} \log \, Z}=
\frac{8 (N^2-1)}{\pi}\int dr \int d\theta \, r^3 \sin^2 \theta \frac{r^2-1-2r^2 \log r}{(1-r^2)^2} \frac{{\cal T}_p (1+r^2-2r\cos \theta, r^2)}{(1+r^2-2r\cos \theta)^2}
\label{twotwopp}
\eal

A second relation involves taking four derivatives with respect to mass deformations \cite{Chester:2020vyz, Chester:2022sqb}
\bal
&-\partial_{m_A} \partial_{m_B} \partial_{m_C} \partial_{m_D} {\cal F} \vert_{m=0}=-\partial_{m_A} \partial_{m_B} \partial_{m_C} \partial_{m_D} {\cal F}^{\text{free}} \vert_{m=0} \, +\\
&\sum_r P^{(ABCD)}_r \frac{k^2}{\pi} \int dr \int d\theta \, r^3 \text{sin}^2 \theta
\frac{\bar D_{1111}(U,V) {\cal G}_r^{\text{int}} (U,V)}{U^2}\bigg|_{U=1+r^2-2r\cos \theta, V=r^2}
\label{flavor4}
\eal
where $k$ is the flavor charge of the global current, and the explicit form of $\bar D_{1111}(U,V)$ can be found in appendix C of \cite{Binder:2018yvd}.

In the particular case of ${\cal N}=4$, the flavor symmetry group is SU(2), and the Cartan subalgebra is just u(1), so one can turn a single mass. Moreover, the ${\cal N}=2$ ${\hat {\mathfrak{B}}}_1$ multiplet is contained in the $S_2$ multiplet, so in this particular instance mass derivatives amount to integrated $S_2$ insertions \cite{Chester:2020vyz}.


\subsection{Computing the planar free energy}
Finally, let's sketch the approach we will follow to compute the free energy of these massive theories in the planar limit \cite{Fiol:2020bhf, Fiol:2021icm}. Supersymmetric localization allows to write the partition function of a ${\cal N}=2$ SYM theory as the following matrix model \cite{Pestun:2007rz}:
\bal
Z=\int da \,\, e^{-\frac{8\pi^2}{g_{YM}^2} \Tr(a^2)} \mathcal{Z}_{1-loop} (a) |\mathcal{Z}_{inst}(a,\tau)|^2
\label{zpestun}
\eal
where $\mathcal{Z}_{inst}$ is the instanton contribution, that can be set to 1 in the large N limit, and the 1-loop determinant depends on the gauge group and the matter content:
\bal
\mathcal{Z}_{1-loop}=\frac{\prod_\alpha H(i\alpha\cdot a)}
{\prod_R \prod_{\omega_R} H(i\omega_R\cdot a)}
\label{z1loop}
\eal
where $H(z)=G(1+z)G(1-z)$ is a product of Barnes G-functions, $\alpha$ are the roots of the adjoint representation of the gauge group, while $\omega_R$ are the weights of the various representations R for the matter multiplets. In this work we will follow an approach advocated in \cite{Billo:2017glv, Billo:2018oog, Billo:2019fbi, Fiol:2018yuc}, where the integral (\ref{zpestun})
is evaluated over the full Lie algebra rather than the usual restriction to a Cartan subalgebra. In all the works considered up to the current one, this results in a rewriting of $\mathcal{Z}_{1-loop}$ as the exponential $\mathcal{Z}_{1-loop}=e^{S_{int}}$ of an effective action with infinitely many single and double trace terms \cite{Billo:2019fbi, Fiol:2020bhf},
\bal
\mathcal{Z}_{1-loop}=e^{S_{\text{int}}}=e^{N\sum_i c_i \text{Tr } a^i+\sum_{i,j} c_{ij}\text{Tr }a^i \text{Tr }a^j}
\label{traceaction}
\eal
In terms of $S_{int}$, the free energy reads
\bal
\cF=-\log Z=-\sum\limits_{m=1}^{\infty} \frac{(-1)^{m+1}}{m} \left( \sum\limits_{k=1}^{\infty} \frac{1}{k!} \langle S_{int}^k\rangle \right)^m
\label{freeenergy}
\eal
so in particular, the planar free energy and the planar n-point functions can be written in terms of planar connected n-point functions of the Gaussian matrix model. For operators of even dimension, these planar connected Gaussian n-point functions are known \cite{tutte, Gopakumar:2012ny}
\bal
\langle \text{Tr } a^{2k_1} \text{Tr }a^{2k_2}\dots \text{Tr }a^{2k_n}\rangle_c =\frac{(d-1)!}{(d-n+2)!} 
\prod_{i=1}^n \frac{(2k_i)!}{(k_i-1)! k_i!} 
\left(\frac{\lambda}{16\pi^2}\right)^d N^{2-n}
\label{gopapius}
\eal
with $d=\sum_{i=1}^n k_i$, so the remaining task is to characterize the terms in (\ref{freeenergy}) that survive in the planar limit. Eventually, one obtains purely combinatorial expressions for the planar free energy and planar n-point functions, as sums over tree graphs \cite{Fiol:2020bhf, Fiol:2021icm}.

As we will see, ${\cal Z}_{1-loop}$ for the most general massive deformation of ${\cal N}=2$ SQCD can't be rewritten as an action involving single and double traces as in (\ref{traceaction}), so we will have to extend the procedure sketched here.

\section{Planar limit of ${\cal N}=4$ integrated correlators}
As a first application, we are going to compute certain fourth derivatives of the planar free energy of a deformation of ${\cal N}=4$ SU(N) SYM on $S^4$. By the relations (\ref{twotwopp}) and (\ref{flavor4}), these are equal to various integrated 4-point functions of ${\cal N}=4$ 1/2 BPS chiral primaries. To be specific, the deformations correspond to adding a mass to the ${\cal N}=2$ hypermultiplet so the theory turns into ${\cal N}=2^*$, plus  further deformations involving ${\cal N}=2$ chiral multiplets. The full action is $S_{{\cal N}=4}+S_m+\sum_p (S_{\tau_p}+S_{\bar \tau_p})$.

For the $\cN=2^*$ theory, the 1-loop determinant (\ref{z1loop}) is
\bal
\mathcal{Z}_{1-loop}=\frac{\prod\limits_{u<v}H(ia_u -ia_v)^2}{\prod\limits_{u<v} H(ia_u -ia_v -M)H(ia_u -ia_v +M)}.
\eal
while the effective action (\ref{traceaction}) is \cite{Fiol:2021jsc}
\bal
S_{int}=&-(N^2-N) \text{ln }H(M)+ 2N \sum\limits_{j=1}^{\infty} \sum\limits_{n=j+1}^{\infty} \frac{\zeta_{2n-1}}{n} \begin{pmatrix}
2n\\2j
\end{pmatrix} (-1)^j M^{2n-2j} \Tr a^{2j}\\
&+ \sum\limits_{i,j=1}^{\infty} \begin{pmatrix}
2i+2j\\ 2i
\end{pmatrix} (-1)^{i+j} \sum\limits_{n=i+j+1} \frac{\zeta_{2n-1}}{n} \begin{pmatrix}
2n\\ 2i+2j
\end{pmatrix} M^{2n-2i-2j} \Tr a^{2i} \Tr a^{2j}\\
& +\sum\limits_{i,j=1}^{\infty} \begin{pmatrix}
2i+2j\\ 2i
\end{pmatrix} (-1)^{i+j} \sum\limits_{n=i+j+2} \frac{\zeta_{2n-1}}{n} \begin{pmatrix}
2n\\ 2i+2j+2
\end{pmatrix} M^{2n-2i-2j-2} \Tr a^{2i+1} \Tr a^{2j+1}
\label{sintn2star}
\eal
We now apply the approach explained in the previous section to compute the planar limit of the free energy on $S^4$, as all-order power series in $\lambda$ around $\lambda=0$. In this particular case, the generic approach sketched above is particularly simple, because we only need to keep terms of $S_{int}$ in (\ref{sintn2star}) that contribute up to $M^4$ to the free energy, so up to the required order
\bal
{\cal F}\vert_{M^4}=-\langle S_{M^2}\rangle -\langle S_{M^4}\rangle-\frac{1}{2}\langle S_{M^2}^2\rangle+\frac{1}{2}\langle S_{M^2}\rangle^2
\eal

\subsection{$ \partial_{\tau_p} \partial_{\bar{\tau_p}} \partial_{m}^2 \log Z$}
As displayed in (\ref{twotwopp}), an integrated version of the ${\cal N}=4$ 4-point function $\langle S_2 S_2 S_p S_p\rangle$ can be obtained from the free energy of the deformation of ${\cal N}=4$ SU(N) SYM just described. We will present the computation for $p=2k$ even, but previous experience with a very similar computation \cite{Fiol:2021icm} immediately suggests the generalization to arbitrary $p$. 

In the planar limit, operators with odd powers in (\ref{sintn2star}) don't contribute, so we effectively need to just consider
\bal
&S_{M^2}=\sum_{i+j\geq 1}^\infty \zeta_{2i+2j+1}(-1)^{i+j} (2i+2j+1){2i+2j \choose 2i} \text{Tr } a^{2i} \text{Tr } a^{2j}
\eal
where we have not written a constant term that doesn't contribute to the fourth derivative of the free energy. We have to add to this action couplings to chiral and antichiral ${\cal N}=2$ operators.

This model falls in the class of models whose planar limit was solved in \cite{Fiol:2021icm}. In particular, it was found in \cite{Fiol:2021icm} that the two-point function of operators with even dimension on $S^4$ is 
\bal
\langle \Omega_{2i} \Omega_{2j}\rangle=f(i)f(j) 
\left(\frac{1}{i+j}-2\sum_{p+q\geq 1}^\infty
c_{2p,2q}f(p)f(q)\left(\frac{1}{p(p+1)}+\frac{1}{(p+i)(q+j)}\right)\right)
\label{twopointS4}
\eal
where $\tilde \lambda =\frac{\lambda}{16\pi^2}$, 
\bal
f(i)=\frac{(2i)!}{i! (i-1)!}\tilde \lambda^i
\eal
and in the case at hand
\bal
c_{2i,2j}=(-1)^{i+j} \zeta_{2i+2j+1}(2i+2j+1) {2i+2j \choose 2i}
\eal

We are ultimately interested in correlators on $\mathbb{R}^4$. The relation among these two sets of operators is subtle, due to operator mixing \cite{Gerchkovitz:2016gxx}. The way to undo this mixing proposed in  \cite{Gerchkovitz:2016gxx} is to apply a Gram-Schmidt orthogonalization procedure. A systematic way to implement this procedure for a very similar problem was presented recently in \cite{Fiol:2022vvv}. The case at hand is even simpler, because since we want to compute $\partial_{\tau_p} \partial_{\bar{ \tau_p}}\partial_{m^2} {\cal F} |_{m^2=0}$, we can truncate the free energy and keep only terms that go like $M^2$. Due to the form of the action (every $\zeta$ goes with an $M^2$), this implies that the surviving terms have maximal transcendentality. Giving names to the two matrices in (\ref{twopointS4}),
\bal
A_{ij}=\frac{1}{i+j}
\eal
\bal
B=B_{ind}+B_{dep}=-2\sum_{p+q\geq 1}^\infty
c_{2p,2q}f(p)f(q)\left(\frac{1}{p(p+1)}+\frac{1}{(p+i)(q+j)}\right)
\label{bindddep}
\eal
then, the quantity $l_p$ defined in (\ref{twotwopp}) is given by
\bal
l_{p} =2 \frac{| \mathbb{I}+A_k^{-1}B|}{|\mathbb{I}+A_{k-1}^1 B|}\,\,
\bigg |_{M^2} =2\, \text{tr } \Delta B
\label{tracedeltab}
\eal
with
\bal
\Delta =A_k^{-1}-A_{k-1}^{-1}
\eal
In \cite{Fiol:2022vvv} the analog of (\ref{tracedeltab}) captured the terms of maximal transcendentality. Now, as remarked above, the terms of maximal transcendentality are the only ones that contribute to $l_p$, so (\ref{tracedeltab}) is exact. As explained in \cite{Fiol:2022vvv}, the matrix A is of Cauchy type, so it is possible to write explicitly the components of its inverse. To compute the trace in (\ref{tracedeltab}) we split it into two parts, for $B_{ind}$ and $B_{dep}$ defined in (\ref{bindddep}). The details of the computations are very similar to the ones in \cite{Fiol:2022vvv}, so we will just present the final results and refer the interested reader to \cite{Fiol:2022vvv} for details of the derivation. We find
\bal
\text{tr }\Delta B_{ind}=-4k \sum_{n=1}^\infty \zeta_{2n+1} (-\tilde \lambda)^n (2n+1) {2n \choose n}{2n \choose n+1}
\eal
while
\bal
\text{tr }\Delta B_{dep}=-4k \sum_{n=1}^\infty \zeta_{2n+1} (-\tilde \lambda)^n (2n+1) {2n \choose n}{2n \choose n+2k}
\eal
Plugging these results into (\ref{tracedeltab}) we find
\bal
l_p= -8k \sum_{n=1}^\infty \zeta_{2n+1} (-\tilde \lambda)^n (2n+1) {2n \choose n}
\left[{2n \choose n+2k}+{2n\choose n+1}\right]
\eal
This is for $p=2k$ even. For arbitrary $p$, based on the results in \cite{Fiol:2021icm}, we expect the answer to be
\bal
l_p
= -4p \sum_{n=1}^\infty \zeta_{2n+1} (-\tilde \lambda)^n (2n+1) {2n \choose n}
\left[(-1)^p{2n \choose n+p}+{2n\choose n+1}\right]
\label{thelp}
\eal
It is worth noticing that the answer we have found, eq. (\ref{thelp}), is very similar to the terms with maximal transcendentality of the planar extremal 2-point functions of ${\cal N}=2$ chiral correlators \cite{Fiol:2021icm, Fiol:2022vvv}, 
\bal
\langle O_{2k} \bar O_{2k}\rangle =2k \tilde \lambda^{2k}
\left[1-4k \sum_{n=2}^\infty \frac{\zeta_{2n-1}(-\tilde \lambda)^n}{n} {2n \choose n}
\left[{2n \choose n+2k}+{2n \choose n+1}-n\right] +\dots \right]
\eal
There are a couple of checks we can perform on our result (\ref{thelp}). First, for $p=2$, the perturbative series around $\lambda=0$ was computed in \cite{Dorigoni:2021guq}, and our result reproduces theirs. Second, using
\bal
J_p(\sqrt{x})^2 =(-1)^p \sum_{n=1}^\infty {2n \choose n} \left(-\frac{x}{4}\right)^n \frac{1}{(n-p)! (n+p)!}
\label{besselseries}
\eal
and
\bal
\int_0^\infty dw\, \, \frac{w^{2n+1}}{\sinh^2 \, w} = \frac{(2n+1)! \zeta_{2n+1}}{4^n}
\label{zetaint}
\eal
we can rewrite (\ref{thelp}) as 
\bal
l_p=-4p \int_0^\infty d\omega \, \frac{\omega}{\sinh^2 \omega} 
\left[J_p\left(\frac{\sqrt{\lambda}\omega}{\pi}\right)^2-
J_1\left(\frac{\sqrt{\lambda}\omega}{\pi}\right)^2\right]  
\eal 
which is precisely the result obtained in \cite{Binder:2019jwn}, see their equation (3.48). It is worth to keep in mind that the derivation in \cite{Binder:2019jwn} assumed $\lambda$ to be large, while our computation is a perturbative expansion around $\lambda=0$. 

As observed already in \cite{Dorigoni:2021guq} for $p=2$, the perturbative series in (\ref{thelp}) all have radius of convergence $|\lambda|=\pi^2$, for any $p$. This value appears to be universal for the planar limit of all quantities that can be derived from ${\cal N}=2$ supersymmetric localization, see \cite{Fiol:2021jsc} and references therein.

\subsection{$\partial_{m}^4 \log Z$}
It was argued in \cite{Chester:2020dja} that a second integrated version of $\langle S_2(x_1)S_2(x_2)S_2(x_3)S_2(x_4)\rangle$ can be obtained by computing $\partial_{m}^4 \log Z$ of the partition function of ${\cal N}=2^*$ on $S^4$, see eq. (\ref{flavor4}). We now proceed to compute this fourth derivative in the planar limit.

To calculate $\partial_{m}^4 \log Z=-\partial_{m}^4 \cF$, we have to consider the terms quartic in $M$ in \eqref{freeenergy}, which include the contributions from $\langle S_{int}\rangle$ and $\langle S_{int}\rangle^2-\langle S_{int}^2\rangle$. Explicitly,
\bal
-\cF \vert_{M^4} =\langle S_{M^4} \rangle +\frac{1}{2} \langle S_{M^2}^2\rangle-\frac{1}{2}\langle S_{M^2}\rangle^2
\label{freeftom4}
\eal
We first look at $\langle S_{M^4} \rangle$ in the planar limit. We need to take into account that the constant term in $S_{int}$, eq. (\ref{sintn2star}), depends on $M$, so it contributes to the result
\bal
\langle S_{int} \rangle \vert_{M^4}= & N^2 \frac{\zeta_3}{2}M^4+
2N \sum\limits_{j=1}^{\infty} \frac{\zeta_{2j+3}}{j+2} \begin{pmatrix}
    2j+4\\2j
\end{pmatrix} (-1)^j M^4 \langle\Tr a^{2j}\rangle\\
&+\sum\limits_{i,j=1}^{\infty} \begin{pmatrix}
    2i+2j\\2i
\end{pmatrix} (-1)^{i+j} \frac{\zeta_{2i+2j+3}}{i+j+2} \begin{pmatrix}
    2i+2j+4\\2i+2j
\end{pmatrix} M^4 \langle \Tr a^{2i} \Tr a^{2j} \rangle\\
=
&\sum\limits_{i,j=0}^{\infty} \begin{pmatrix}
    2i+2j\\2i
\end{pmatrix} (-1)^{i+j} \frac{\zeta_{2i+2j+3}}{i+j+2} \begin{pmatrix}
    2i+2j+4\\2i+2j
\end{pmatrix} M^4 \langle \Tr a^{2i} \Tr a^{2j} \rangle
\eal
where the last equality holds in the planar limit. These sums are readily evaluated, yielding
\bal
\langle S_{int} \rangle \vert_{M^4} = 
N^2 M^4 \frac{1}{12} \sum_{n=1}^\infty \zeta_{2n+1} (-\tilde \lambda)^{n-1}
(2n+1) {2n \choose n} {2n \choose n+1}
\eal

As for the $\langle S^2\rangle -\langle S \rangle^2$ term in (\ref{freeftom4}), the contribution in the planar limit is
\bal
& \frac{1}{2}(\langle S^2\rangle -\langle S \rangle^2)= 
 2\sum_{i+j\geq 1}^\infty \sum_{k+l\geq 1}^\infty \zeta_{2i+2j+1} \zeta_{2k+2l+1} (-1)^{i+j+k+l} \\
& (2i+2j+1){2i+2j \choose 2i} (2k+2l+1) {2k+2l \choose 2k} \langle \text{Tr }a^{2i} \rangle \langle \text{Tr } a^{2j} \text{Tr } a^{2k} \rangle_c \langle \text{Tr } a^{2l} \rangle 
\eal
which can be visualized as coming from the single tree graph with three vertices. It is possible to carry out some of the intermediate sums, and we arrive at
\bal
& \frac{1}{2}(\langle S^2\rangle -\langle S \rangle^2)= \\
& 2 \sum_{m,n=1}^\infty \zeta_{2m+1}\zeta_{2n+1} (-\tilde \lambda)^{m+n}
\frac{(2m+1)!}{m!\, m!} \frac{(2n+1)!}{n!\, n!} \sum_{i=1}^m {m \choose i}{m \choose i-1} \sum_{j=1}^n {n \choose j}{n \choose j-1} \frac{1}{i+j}
\label{s2minuss2}
\eal
Thus, our result for the fourth mass derivative of the planar free energy is
\bal 
& -\partial_m^4 {\cal F}_0 |_{m=0}= 2 \sum_{n=1}^\infty \zeta_{2n+1} (-\tilde \lambda)^{n-1} (2n+1) {2n \choose n} {2n \choose n+1} + \\
& 48 \sum_{m,n=1}^\infty \zeta_{2m+1}\zeta_{2n+1} (-\tilde \lambda)^{m+n}
\frac{(2m+1)!}{m!\, m!} \frac{(2n+1)!}{n!\, n!} \sum_{i=1}^m {m \choose i}{m \choose i-1} \sum_{j=1}^n {n \choose j}{n \choose j-1} \frac{1}{i+j}
\label{ourf4m}
\eal
We can compare this perturbative series with the result found in \cite{Chester:2020dja}.
Their equation (A.24) is
\bal
&-\partial_m^4 {\cal F}_0^{pert}\vert_{m=0} = \int_0^\infty dx \, 
\frac{32 x \pi^2 J_1(\frac{\sqrt{\lambda}x}{\pi})^2}{\lambda \sinh^2 x} \\
& +\int_0^\infty dx \, \int_0^\infty dy \, \frac{96 \pi xy J_1(\frac{\sqrt{\lambda}x}{\pi}) J_1(\frac{\sqrt{\lambda}y}{\pi})}{\sqrt{\lambda} \sinh^2 x \, \sinh^2 y \, (y^2-x^2)}\left[x J_0(\frac{\sqrt{\lambda}x}{\pi}) J_1(\frac{\sqrt{\lambda}y}{\pi})-y J_0(\frac{\sqrt{\lambda}y}{\pi}) J_0(\frac{\sqrt{\lambda}x}{\pi}) \right]
\label{d4mplanar}
\eal
Using (\ref{besselseries}) and (\ref{zetaint}) we can rewrite the first line of (\ref{ourf4m}) in integral form, and  we recover exactly the first line in (\ref{d4mplanar}). As for the sum with two values of $\zeta$ in (\ref{ourf4m}), it ought to be the perturbative expansion of the double integral in (\ref{d4mplanar}). In order to check this, we have derived the perturbative series of the double integral in (\ref{d4mplanar}) - see appendix \ref{recursion} for details - and obtained
\bal
\label{expansion}
&\int_0^{\infty} dw_1 \int_0^{\infty} dw_2 \frac{w_1 w_2}{(w_1^2 -w_2^2)\sinh^2 w_1 \sinh^2 w_2} \\
&\left[w_2 J_1(\frac{\sqrt{\lambda} w_1}{\pi})^2 J_1 (\frac{\sqrt{\lambda} w_2} {\pi}) J_0(\frac{\sqrt{\lambda} w_2}{\pi})
-w_1 J_1(\frac{\sqrt{\lambda} w_2}{\pi})^2 J_1 (\frac{\sqrt{\lambda} w_1} {\pi}) J_0(\frac{\sqrt{\lambda} w_1}{\pi})\right] =\\
& \sum\limits_{m,n=1}^{\infty} f_{mn} (-\frac{\lambda}{16 \pi^2})^{m+n} (2m+1)
! (2n+1)! \zeta_{2m+1} \zeta_{2n+1}
\eal

with 
\bal
\label{fmn}
& f_{mn}=\frac{1}{2}\frac{(2m+1)! (2n)!}{(m+2)! (m+1)! m!^2 (+1)! n!^2 (n-1)!} \\
& \left[ (m-n-1) \,\, _5F_4 (1,m+\frac{3}{2},-n-1,1-n,-n;m+1,m+2,m+3,\frac{1}{2}-n;1)\, + \right. \\
& \left. 2 \,\, _5F_4 (2,m+\frac{3}{2},-n-1,1-n,-n;m+1,m+2,m+3,\frac{1}{2}-n;1) \right] 
\eal

These coefficients are actually symmetric in $m,n$, although this is not manifest from the expression above. While we haven't proved that these coefficients coincide with our result (\ref{ourf4m}), as explained in appendix \ref{recursion}, we have checked that they agree for $m,n=1,\dots, 100$. We are thus rather confident that our result (\ref{ourf4m}) agrees with the result (\ref{d4mplanar}) obtained in \cite{Chester:2020dja}, and can be fed into the ${\cal N}=4$ version of (\ref{flavor4}) to obtain an integrated relation for $\langle S_2(x_1)S_2(x_2)S_2(x_3)S_2(x_4)\rangle$.

\section{Planar limit of ${\cal N}=2$ integrated correlators}
As a second application, we will now study integrated 4-point functions of a genuine ${\cal N}=2$ theory, SU(N) with $N_f=2N$ hypermultiplets in the fundamental representation, the so-called ${\cal N}=2$ SU(N) SQCD. To do so, according to the discussion in section 2, we consider its massive deformation, by adding masses to the hypermultiplets. This massive deformation was studied using supersymmetric localization in \cite{Russo:2013kea}. In this case, the $\mathcal{Z}_{1-loop}$ in eq. (\ref{z1loop}) is

\bal
\mathcal{Z}_{1-loop}=\frac{\prod\limits_{u<v}H(ia_u -ia_v)^2}{\prod\limits_{u=1}^N H(ia_u +M_u)^N H(ia_u -M_u)^N}.
\eal
Taking the logarithm and keeping terms with up to $M^4$ we find
\bal
S_{int}\vert_{M^4}=  N \sum_{n=2}^\infty \left[c_n \sum_{u=1}^N M_u^2 a_u^{2n-2} + d_n \sum_{u=1}^N M_u^4 \, a_u^{2n-4} 
\right] + \sum_{n=2}^\infty \sum_{k=1}^{n-1} c_{n-k,k} \text{tr }a^{2n-2k} \text{tr }a^{2k}
\label{sintsqcd}
\eal
where the various coefficients are
\bal   
c_n=-2\frac{\zeta_{2n-1}}{n}(-1)^n {2n \choose 2}
\hspace{.5cm}
d_n=2\frac{\zeta_{2n-1}}{n}(-1)^n {2n \choose 4}
\hspace{.5cm}
c_{ij}=-\frac{\zeta_{2i+2j-1}}{i+j}(-1)^{i+j} {2i+2j\choose 2i}
\eal  
In (\ref{sintsqcd}) we have omitted a constant term and a term with traces of operators with odd dimensions since they don't contribute in the planar limit. The corresponding free energy is an even function of each of the masses, and it must be symmetric under their permutation, so the only two possible terms in the planar free energy that are quartic in the masses are of the form 
\bal
{\cal F}\vert_{M^4}= \alpha \, \sum_{u=1}^N M_u^4 +\beta \sum_{u\neq v} M_u^2 M_v^2
\label{freetwoterms}
\eal
and there are thus two independent fourth-derivatives to consider
\bal
\frac{1}{4!}\partial^4_{M_u} {\cal F} \vert_{M_u=0} = \alpha , \hspace{1cm}
\frac{1}{8}\partial^2_{M_u} \partial^2_{M_v} {\cal F} \vert_{M_u=0} = \beta \hspace{.5cm} u\neq v
\eal
Our goal is this section is to obtain $\alpha $ and $\beta $ in the planar limit. These can be then used in (\ref{flavor4}) to obtain identities involving integrals of ${\cal G}_r(U,V)$.

It will prove convenient to write ${\cal F}$ in the following alternative way
\bal
{\cal F}= N \alpha_1 (\lambda) \, \sum_{u=1}^N M_u^4 +\alpha_2 (\lambda) \sum_{u\neq v} M_u^2 M_v^2 + N^2 \alpha_3(\lambda) \, \left( \sum_{u=1}^N M_u^4-\frac{1}{N-1} \sum_{u\neq v} M_u^2 M_v^2\right)
\label{3alphas}
\eal
Note that the last term in (\ref{3alphas}) vanishes when all the masses are the same, $M_u=M$. It is immediate to write $\alpha(\lambda), \beta(\lambda)$ in terms of $\alpha_{1,2,3}(\lambda)$
\bal 
\alpha(\lambda)=N \alpha_1(\lambda)+ N^2 \alpha_3(\lambda)\hspace{1cm} \beta(\lambda)=\alpha_2(\lambda)-\frac{N^2}{N-1}\alpha_3(\lambda)
\label{alphabeta}
\eal
The reasons why (\ref{3alphas}) is a convenient way to present ${\cal F}$ are the following. First, it allows to make explicit the large N scaling of the different terms, and the dominance of $\alpha_3$ in the large N limit; we will argue soon that the coefficient to the last term in (\ref{3alphas}) admits a $1/N^2$ expansion, so the first subleading correction is of order $N^0$. Thus it is consistent to keep $\alpha_1(\lambda),\alpha_2(\lambda)$ as the first subleading corrections in (\ref{alphabeta}). Moreover, the effective action (\ref{sintsqcd}), for arbitrary masses, can't be written in terms of traces of powers of $a$. Therefore it falls outside the family of matrix models discussed at the end of section 2. Nevertheless, we will argue that $\alpha_1(\lambda)$ and $\alpha_2(\lambda)$ in (\ref{3alphas}) can be obtained by setting all masses equal, $M_u=M$. When all masses are equal, (\ref{sintsqcd}) can be written as a matrix model of single and double traces, and thus $\alpha_{1,2}$ admit closed expressions given by sum over tree graphs. On the other hand, $\alpha_3(\lambda)$ can't be derived with the same techniques. We will nevertheless present a procedure to compute it order by order in transcendentality.

Let's now consider all the possible insertions from (\ref{sintsqcd}) of order $M^4$ that can appear in ${\cal F}$. A first type of contribution to ${\cal F}$ is a single $N d_n M_u^4 a_u^{2n-4}$ insertion from (\ref{sintsqcd}). These are the only contributions to $\alpha_1$ in (\ref{3alphas}). When taking expected values, we can use
\bal
\langle a_u^{2n-4} \, *\rangle = \frac{1}{N} \langle \text{tr } a^{2n-4} \, *\rangle
\eal
where here and in all the similar equations below, $*$ denotes an arbitrary product of traces of powers of $a$. As a consequence, the contribution of all the single $M_u^4$ insertions can be obtained by trading in (\ref{sintsqcd})
\bal
\sum_{u=1}^N M_u^4 a_u^{2n-4} \rightarrow \frac{1}{N}\sum_{u=1}^N M_u^4 \text{tr }a^{2n-4}
\eal
The only other possible way to have order $M^4$ terms coming from (\ref{sintsqcd}) in ${\cal F}$ is having two $N c_m M_u^2 a_u^{2m-2}$ insertions, coming from $S_{int}^2$ in (\ref{freeenergy}). When taking expected values, these two insertions can appear either in the same correlator, or in two different correlators. In either case, we can write the outcome in terms of expected values of traces. When both insertions are in the same correlator we have
\bal
&\langle \sum_{u,v} M_u^2 M_v^2 a_u^m a_v^n *\rangle =\\
& \left( \sum_{u=1}^N M_u^4-\frac{1}{N-1} \sum_{u\neq v} M_u^2 M_v^2\right) \langle \frac{1}{N}\text{tr }a^{m+n} *\rangle + \frac{1}{N(N-1)} \sum_{u\neq v} M_u^2 M_v^2 \langle \text{tr }a^m \text{tr }a^n *\rangle
\label{atotraces1}
\eal  
while in the case of the insertions being in different correlators we obtain  
\bal   
&\langle \sum_u M_u^2 a_u^m *_1\rangle \langle \sum _v M_v^2 a_v^n *_2\rangle =
 \left( \sum_{u=1}^N M_u^4-\frac{1}{N-1} \sum_{u\neq v} M_u^2 M_v^2\right)
\langle \frac{1}{N} \text{tr }a^m *_1\rangle \langle \frac{1}{N} \text{tr }a^n *_2 \rangle
+ \\
& \frac{1}{N(N-1)} \sum_{u\neq v} M_u^2 M_v^2 \langle \text{tr }a^m *_1\rangle \langle \text{tr }a^n *_2\rangle
\label{atotraces2}
\eal
Both (\ref{atotraces1}) and (\ref{atotraces2}) are valid at finite N. Note that in both cases, the last terms in the right hand sides of (\ref{atotraces1}) and (\ref{atotraces2}) are what we would obtain in the case when all the masses are equal $M_u=M$. These terms will contribute to $\alpha_2$ in (\ref{3alphas}). The other two terms in  (\ref{atotraces1}) and (\ref{atotraces2}) contribute to $\alpha_3$.
The outcome of this analysis is that $\alpha_1$ and $\alpha_2$ can be readily obtained by applying the methods of \cite{Fiol:2020bhf, Fiol:2021icm} to the effective action (\ref{sintsqcd}) in the case when all masses are equal,
\bal
 S_{int}\vert_{M_u=M}= N \sum_{n=2}^\infty \left[M^2 c_n \text{tr }a^{2n-2} +M^4 d_n \text{tr }a^{2n-4} 
\right] 
 + \sum_{n=2}^\infty \sum_{k=1}^{n-1} c_{n-k,k} \text{tr }a^{2n-2k} \text{tr }a^{2k}
\label{equalmass}
\eal
Note that the single trace terms in (\ref{equalmass}) have an explicit factor of N in form of it, so they have the right scaling to contribute to the planar limit. The matrix model with (\ref{equalmass}) as potential falls into the class of models solved in \cite{Fiol:2021icm}. The planar free energy is given by a sum over tree graphs with an arbitrary number of edges. We will now proceed to compute $\alpha_1(\lambda)$ and $\alpha_2(\lambda)$ and return to $\alpha_3$ later. Let's discuss in turns the contributions to $\alpha_1$ and $\alpha_2$. The first corresponds to insertions of $M^4 \,d_n\, \text{tr a}^{2n-4}$,
\bal      
\alpha_1 (\lambda) = -\sum_{m=1}^\infty \frac{1}{(m-1)!} \sum_{n=2}^\infty d_n \sum_{i_1,j_1,\dots i_{m-1},j_{m-1}} c_{i_1j_1}\dots c_{i_{m-1}j_{m-1}} \sum_{\text{trees}_1} \prod_{i=1}^m V_i
\label{alpha1trees}
\eal        
where the sum $\text{trees}_1$ is over rooted directed trees with $m-1$ labeled edges\footnote{A rooted tree is a tree with a distinguished vertex. A directed tree is one where the edges have directions. A detailed discussion of the relevant notions of graph theory that appear in these types of computations can be found in \cite{Fiol:2020bhf}.}. The $V_i$ is the planar connected correlator (\ref{gopapius}) associated with the i-$th$ vertex of the tree, except for the power of N: it contains $\text{tr } a^{i_s}$ if the directed edge $s$ leaves the $i$ vertex, $\text{tr }a^{j_s}$ if the directed edge $s$ arrives at the $i$ vertex, and $\text{tr }a^{2n-4}$ if the vertex is the rooted one. To illustrate (\ref{alpha1trees}), let's write down explicitly the terms with leading and subleading transcendentality. They are encoded in tree graphs with zero and one edge respectively. In the first case, the single vertex corresponds to $\langle \text{tr } a^{2n-4} \rangle$; in the second case, this insertion can be in any of the two vertices connected by the edge:
\bal
& \alpha_1 (\lambda) =-\, \sum_{n=2}^\infty d_n \langle \text{tr }a^{2n-4}\rangle - 2\sum_{n=2}^\infty d_n
\sum_{i,j=1}^\infty c_{ij}  \langle \text{tr }a^{2i}\rangle \langle \text{tr }a^{2j} \text{tr }a^{2n-4}\rangle_c = \\
& -2\, \sum_{n=2}^\infty \frac{\zeta_{2n-1}}{n} {2n\choose 4} (-\tilde \lambda)^{n-2} \left[ {\cal C}_{n-2}-2\frac{(2n-4)!}{(n-3)!(n-2)!}\sum_{m=2}^\infty \frac{\zeta_{2m-1} (-\tilde \lambda)^m}{m} {2m \choose m}  \right. \\
&\left. \sum_{i=1}^{m-1}{m\choose i}{m \choose i-1}\frac{1}{i+n-2}  \right]
\eal
where ${\cal C}_{n-2}$ is a Catalan number. Turning now to $\alpha_2(\lambda)$, it receives contributions from two insertions of the type $M^2 c_n \, \text{tr } a^{2n-2}$. They can appear in the same vertex or in different vertices; we denote these two possibilities by $\alpha_{2a}$ and $\alpha_{2b}$, so $\alpha_2=\alpha_{2a}+\alpha_{2b}$. When they appear in the same vertex, 
\bal   
\alpha_{2a}(\lambda)= -\frac{1}{2}\sum_{m=0}^\infty \frac{1}{m!} \sum_{n_1,n_2} c_{n_1} c_{n_2} \sum_{i_1,j_1,\dots i_{m},j_{m}} c_{i_1 j_1}\dots c_{i_m j_m} \sum_{\text{trees}_1} \prod_{i=1}^{m+1} V_i 
\eal
where both insertions are in the same vertex. The leading and first subleading terms in transcendentality are
\bal 
& \alpha_{2a}(\lambda)=-\frac{1}{2} \sum_{m,n=2}^\infty c_m c_n \left[ \langle \text{tr }a^{2m-2} \text{tr }a^{2n-2} \rangle_c+2\sum_{i,j=1}^\infty c_{ij} \langle \frac{1}{N} \text{tr }a^{2i} \rangle \langle \text{tr }a^{2j} \text{tr }a^{2m-2} \text{tr }a^{2n-2}\rangle_c \right]=\\
& -2\, \sum_{m,n=2}^\infty \zeta_{2m-1} \zeta_{2n-1} (-\tilde \lambda)^{m+n-2} \frac{(2m-1)!}{(m-2)! (m-1)!} \frac{(2n-1)!}{(n-2)! (n-1)!} \\
& \left[ \frac{1}{m+n-2}-2\sum_{r=2}^\infty \zeta_{2r-1} (-\tilde \lambda)^r {2r \choose r} ({\cal C}_r-1)  \right]
\eal
with ${\cal C}_r$ a Catalan number. When the two insertions take place at {\it different} vertices we find
\bal   
\alpha_{2b}(\lambda)=-\frac{1}{2}\sum_{m=0}^\infty \frac{1}{m!} \sum_{n_1,n_2} c_{n_1} c_{n_2} \sum_{i_1,j_1,\dots i_m,j_m} c_{i_1 j_1}\dots c_{i_m j_m} \sum_{\text{trees}_2} \prod_{i=1}^{m+1} V_i 
\eal   
where the sum is over doubly rooted directed trees with $m-1$ labeled edges. The two marked vertices have to be different. The $V_i$ are as before, with now the insertion of $\text{tr } a^{2n_i-2}$ if the vertex is rooted. The term of leading transcendentality is
\bal
{\alpha}_{2b}(\lambda)=-4 \sum_{m,n=2}^\infty \zeta_{2m-1} \zeta_{2n-1} (-\tilde \lambda)^{m+n-2} \frac{(2m-1)!}{(m-2)! (m-1)!} \frac{(2n-1)!}{(n-2)! (n-1)!}\\
\sum_{i,j=1}^\infty \frac{\zeta_{2i+2j-1}}{i+j}(-\tilde \lambda)^{i+j} {2i+2j \choose 2i} \frac{(2i)! (2j)!}{(i-1)! i! (j-1)! j!}\frac{1}{m+i-1}\frac{1}{n+j-1}+\dots
\eal

To conclude, let's explain how to obtain $\alpha_3(\lambda)$. From (\ref{atotraces1}) and (\ref{atotraces2}) we learn that this term captures the failure of the naive substitution $a_u^{2m} \rightarrow \frac{1}{N}\text{tr } a^{2m}$ in giving the correct answer. Then, as we expand (\ref{freeenergy}) with (\ref{sintsqcd}) keeping up to two $M^2$ insertions, on top of the terms that contribute to $\alpha_2$, we apply the following procedure. If the two $M^2$ insertions appear in the same correlator, we add the first term in the right hand side (RHS) of (\ref{atotraces1}). If the two $M^2$ insertions are in different correlators, we add the first term in the RHS of (\ref{atotraces2}). After this step is completed, we have to rewrite the resulting correlators in terms of connected correlators. Once we accomplish this, we can evaluate the planar connected correlators using (\ref{gopapius}). 
To conclude let's argue that the coefficient of the last term in (\ref{3alphas}) admits a $1/N^2$ expansion. First, as already mentioned, the identities (\ref{atotraces1}) and (\ref{atotraces2}) are valid for finite N. The large N expansion is used in the procedure above only when we evaluate the n-point functions of the Gaussian Hermitian matrix model, and it is well known that these admit a $1/N^2$ expansion, so the same is true for the subleading corrections to $N^2 \alpha_3(\lambda)$. Let's illustrate the procedure by computing the terms in leading and subleading transcendentality for $\alpha_3 (\lambda)$

\bal  
& \alpha_3 (\lambda)= -\frac{1}{2} \sum_{m,n=2}^\infty c_m c_n \bigg\{ \langle \frac{1}{N} \text{tr } a^{2m+2n-4}\rangle -\langle \frac{1}{N} \text{tr }a^{2m-2} \rangle \langle \frac{1}{N} \text{tr } a^{2n-2}\rangle \\
& +\sum_{i,j=1}^\infty c_{ij} \left[ \langle \text{tr }a^{2i}\text{tr }a^{2j} \frac{1}{N} \text{tr }a^{2m+2n-2}\rangle -\langle \text{tr }a^{2i}
\text{tr }a^{2j} \rangle \langle \frac{1}{N}\text{tr }a^{2m+2n-4}\rangle 
\right. \\
& \left. -2 \langle \frac{1}{N} \text{tr }a^{2m-2} \rangle \langle \text{tr }a^{2i}\text{tr }a^{2j} \frac{1}{N}\text{tr }a^{2n-2}\rangle +2\langle \text{tr }a^{2i}\text{tr }a^{2j} \rangle \langle \frac{1}{N}\text{tr }a^{2m-2}\rangle \langle \frac{1}{N}\text{tr }a^{2n-2}\rangle 
\right]\bigg\}
\eal
After rewriting the correlators in terms of connected correlators and applying (\ref{gopapius}) we arrive at
\bal
& \alpha_3 (\lambda)=-2 \sum_{m,n=2}^\infty \zeta_{2m-1} \zeta_{2n-1} (2m-1)(2n-1) (-\tilde \lambda)^{m+n-2} \bigg\{ {\cal C}_{m+n-2}-{\cal C}_{m-1}{\cal C}_{n-1}  \\
& -2 \sum_{i,j=1}^\infty \frac{\zeta_{2i+2j-1}}{i+j} (-\tilde \lambda)^{i+j}
\frac{(2i+2j)!}{(i+1)! i! (j-1)! j!} \left[ \frac{1}{j+m+n-2}\frac{(2m+2n-4)!}{(m+n-3)!(m+n-2)!} \right. \\
& 
\left. \left. -\frac{2}{j+n-1} \frac{(2m-2)!}{m! (m-1)!}\frac{(2n-2)!}{(n-2)!(n-1)!}\right] \right\}
\eal

We have thus managed to compute the planar limit of $\partial^4_{M_u} {\cal F}$ and $\partial^2_{M_u}\partial^2_{M_v} {\cal F}$ to leading an subleading order in transcendality. We also have derived the respective leading 1/N corrections to all order in transcendentality. These derivatives can now be plugged in (\ref{flavor4}) to obtain integrated relations for ${\cal G}_r^{int}(U,V)$.

\acknowledgments
We would like to thank Fernando Alday and Shai Chester for correspondence. The research of BF is supported by  the State Agency for Research of the Spanish Ministry of Science and Innovation through the ``Unit of Excellence Mar\'ia de Maeztu 2020-2023'' award to the Institute of Cosmos Sciences (CEX2019-000918-M) and PID2019-105614GB-C22. ZK is supported by CSC grant No. 201906340174.

\appendix

\section{Perturbative expansions}
\label{recursion}
We want to find the perturbative power series of the second integral in (\ref{d4mplanar}) around $\lambda=0$ and compare it with the one obtained by our methods, eq. (\ref{s2minuss2}) . Define $a\equiv \frac{\sqrt{\lambda}}{\pi}$; the integral to expand is
\bal
\frac{96}{a} \int_0^\infty dx\, \int_0^\infty dy \, \frac{x\, y}{\sinh^2 x \sinh^2 y}\frac{J_1(ax) J_1 (ay)}{y^2-x^2}\left( x J_0(ax) J_1(ay)-yJ_0(ay)J_1(ax)\right)
\label{theintapp}
\eal
We use the perturbative series
\bal
& J_0(x) J_1(x) =\sum_{k=0}^\infty \frac{(-1)^k (2k+1)!}{k!^2 (k+1)!^2}\left(\frac{x}{2} \right)^{2k+1} & \equiv  & \,\,\, \sum_{k=0}^\infty c_k \left(\frac{x}{2} \right)^{2k+1} \\
& J_1(x) J_1(x) =\sum_{k=0}^\infty \frac{(-1)^k (2k+2)!}{k! (k+1)!^2 (k+2)!}\left(\frac{x}{2} \right)^{2k+2} & \equiv & \, \, \, \sum_{k=0}^\infty d_k \left(\frac{x}{2} \right)^{2k+2}
\eal
In terms of these coefficients, the expansion of the numerator in (\ref{theintapp}) is immediate
\bal
J_1(ax) J_1(ay) \left(x J_0(ax) J_1(ay)-yJ_0(ay)J_1(ax) \right) =\frac{2}{a} \sum_{k,l=0}^\infty 
\left(c_k d_l-c_l d_k\right) \left(\frac{ax}{2}\right)^{2k+2} \left(\frac{ay}{2}\right)^{2l+2}
\eal
while the perturbative series we are after is
\bal
\frac{J_1(ax) J_1(ay)}{y^2-x^2} \left(x J_0(ax) J_1(ay)-yJ_0(ay)J_1(ay)\right)=
\frac{a}{2} \sum_{i,j=0}^\infty h_{ij} \left(\frac{ax}{2}\right)^{2i} \left(\frac{ay}{2}\right)^{2j}
\eal
These coefficients satisfy the following relation
\bal
h_{i+1,j}-h_{i,j+1}=c_id_j-c_j d_i
\eal
which is solved by
\bal
h_{ij}=\sum_{k=0}^{j-1} c_k d_{i+j-1-k}-c_{i+j-1-k}d_k
\eal
Carrying out the sum we find,
\bal
f_{ij}\equiv \frac{(-1)^{i+j}}{2}h_{ij}= \frac{1}{4} \frac{(2i+2j+2)!}{(i+j+1)!^4} \sum_{k=0}^{\text{min }(i,j)-1} \frac{ {i+j+1 \choose k} {i+j+1 \choose k+1}^2 {i+j+1 \choose k+2}}{{2i+2j+2 \choose 2k+2}} (i+j-1-2k)
\eal

According to Mathematica, these coefficients can be written as a sum of hypergeometric functions
\bal
& f_{ij}=\frac{1}{2}\frac{(2i+1)! (2j)!}{(i+2)! (i+1)! i!^2 (j+1)! j!^2 (j-1)!} \\
& \left[ (i-j-1) \,\, _5F_4 (1,i+\frac{3}{2},-j-1,1-j,-j;i+1,i+2,i+3,\frac{1}{2}-j;1)\, + \right. \\
& \left. 2 \,\, _5F_4 (2,i+\frac{3}{2},-j-1,1-j,-j;i+1,i+2,i+3,\frac{1}{2}-j;1) \right] 
\eal
Comparing with the perturbative series found by our methods, eq. (\ref{s2minuss2}), their equivalence amounts to the identity
\bal
& \frac{1}{4} \frac{(2i+2j+2)!}{(i+j+1)!^4} \sum_{k=0}^{\text{min }(i,j)-1} \frac{ {i+j+1 \choose k} {i+j+1 \choose k+1}^2 {i+j+1 \choose k+2}}{{2i+2j+2 \choose 2k+2}} (i+j-1-2k) \stackrel{?}{=} \\
&\frac{1}{2}\frac{1}{i!^2 j!^2} \sum_{k=1}^i {i \choose k}{i\choose k-1}\sum_{l=1}^j {j\choose l}{j\choose l-1}\frac{1}{k+l}
\eal
We haven't proved this identity, but we have checked that it holds for $i,j=1,\dots, 100$.

\bibliographystyle{utphys2}
\bibliography{refs}

\providecommand{\href}[2]{#2}\begingroup\raggedright\begin{thebibliography}{10}\setlength{\parskip}{1pt}\setlength{\itemsep}{0pt
  plus 0.3ex}

\bibitem{Dolan:2002zh}
F.~A. Dolan and H.~Osborn, ``{On short and semi-short representations for
  four-dimensional superconformal symmetry},''
  \href{http://dx.doi.org/10.1016/S0003-4916(03)00074-5}{{\em Annals Phys.}
  {\bfseries 307} (2003) 41--89},
  \href{http://arxiv.org/abs/hep-th/0209056}{{\ttfamily arXiv:hep-th/0209056}}.

\bibitem{Baggio:2012rr}
M.~Baggio, J.~de~Boer, and K.~Papadodimas, ``{A non-renormalization theorem for
  chiral primary 3-point functions},''
  \href{http://dx.doi.org/10.1007/JHEP07(2012)137}{{\em JHEP} {\bfseries 07}
  (2012) 137}, \href{http://arxiv.org/abs/1203.1036}{{\ttfamily arXiv:1203.1036
  [hep-th]}}.

\bibitem{Heslop:2022qgf}
P.~Heslop, ``{The SAGEX Review on Scattering Amplitudes, Chapter 8: Half BPS
  correlators},'' \href{http://dx.doi.org/10.1088/1751-8121/ac8c71}{{\em J.
  Phys. A} {\bfseries 55} no.~44, (2022) 443009},
  \href{http://arxiv.org/abs/2203.13019}{{\ttfamily arXiv:2203.13019
  [hep-th]}}.

\bibitem{Dobrev:1985qv}
V.~K. Dobrev and V.~B. Petkova, ``{All Positive Energy Unitary Irreducible
  Representations of Extended Conformal Supersymmetry},''
  \href{http://dx.doi.org/10.1016/0370-2693(85)91073-1}{{\em Phys. Lett. B}
  {\bfseries 162} (1985) 127--132}.

\bibitem{Papadodimas:2009eu}
K.~Papadodimas, ``{Topological Anti-Topological Fusion in Four-Dimensional
  Superconformal Field Theories},''
  \href{http://dx.doi.org/10.1007/JHEP08(2010)118}{{\em JHEP} {\bfseries 08}
  (2010) 118}, \href{http://arxiv.org/abs/0910.4963}{{\ttfamily arXiv:0910.4963
  [hep-th]}}.

\bibitem{Beem:2013sza}
C.~Beem, M.~Lemos, P.~Liendo, W.~Peelaers, L.~Rastelli, and B.~C. van Rees,
  ``{Infinite Chiral Symmetry in Four Dimensions},''
  \href{http://dx.doi.org/10.1007/s00220-014-2272-x}{{\em Commun. Math. Phys.}
  {\bfseries 336} no.~3, (2015) 1359--1433},
  \href{http://arxiv.org/abs/1312.5344}{{\ttfamily arXiv:1312.5344 [hep-th]}}.

\bibitem{Eden:2000qp}
B.~U. Eden, P.~S. Howe, A.~Pickering, E.~Sokatchev, and P.~C. West, ``{Four
  point functions in N=2 superconformal field theories},''
  \href{http://dx.doi.org/10.1016/S0550-3213(00)00218-2}{{\em Nucl. Phys. B}
  {\bfseries 581} (2000) 523--558},
  \href{http://arxiv.org/abs/hep-th/0001138}{{\ttfamily arXiv:hep-th/0001138}}.

\bibitem{Nirschl:2004pa}
M.~Nirschl and H.~Osborn, ``{Superconformal Ward identities and their
  solution},'' \href{http://dx.doi.org/10.1016/j.nuclphysb.2005.01.013}{{\em
  Nucl. Phys. B} {\bfseries 711} (2005) 409--479},
  \href{http://arxiv.org/abs/hep-th/0407060}{{\ttfamily arXiv:hep-th/0407060}}.

\bibitem{Dolan:2004mu}
F.~A. Dolan, L.~Gallot, and E.~Sokatchev, ``{On four-point functions of 1/2-BPS
  operators in general dimensions},''
  \href{http://dx.doi.org/10.1088/1126-6708/2004/09/056}{{\em JHEP} {\bfseries
  09} (2004) 056}, \href{http://arxiv.org/abs/hep-th/0405180}{{\ttfamily
  arXiv:hep-th/0405180}}.

\bibitem{Beem:2014zpa}
C.~Beem, M.~Lemos, P.~Liendo, L.~Rastelli, and B.~C. van Rees, ``{The $
  \mathcal{N}=2 $ superconformal bootstrap},''
  \href{http://dx.doi.org/10.1007/JHEP03(2016)183}{{\em JHEP} {\bfseries 03}
  (2016) 183}, \href{http://arxiv.org/abs/1412.7541}{{\ttfamily arXiv:1412.7541
  [hep-th]}}.

\bibitem{Pestun:2007rz}
V.~Pestun, ``{Localization of gauge theory on a four-sphere and supersymmetric
  Wilson loops},'' \href{http://dx.doi.org/10.1007/s00220-012-1485-0}{{\em
  Commun. Math. Phys.} {\bfseries 313} (2012) 71--129},
  \href{http://arxiv.org/abs/0712.2824}{{\ttfamily arXiv:0712.2824 [hep-th]}}.

\bibitem{Argyres:2015ffa}
P.~Argyres, M.~Lotito, Y.~L\"u, and M.~Martone, ``{Geometric constraints on the
  space of $ \mathcal{N} $ = 2 SCFTs. Part I: physical constraints on relevant
  deformations},'' \href{http://dx.doi.org/10.1007/JHEP02(2018)001}{{\em JHEP}
  {\bfseries 02} (2018) 001}, \href{http://arxiv.org/abs/1505.04814}{{\ttfamily
  arXiv:1505.04814 [hep-th]}}.

\bibitem{Cordova:2016xhm}
C.~Cordova, T.~T. Dumitrescu, and K.~Intriligator, ``{Deformations of
  Superconformal Theories},''
  \href{http://dx.doi.org/10.1007/JHEP11(2016)135}{{\em JHEP} {\bfseries 11}
  (2016) 135}, \href{http://arxiv.org/abs/1602.01217}{{\ttfamily
  arXiv:1602.01217 [hep-th]}}.

\bibitem{Gerchkovitz:2016gxx}
E.~Gerchkovitz, J.~Gomis, N.~Ishtiaque, A.~Karasik, Z.~Komargodski, and S.~S.
  Pufu, ``{Correlation Functions of Coulomb Branch Operators},''
  \href{http://dx.doi.org/10.1007/JHEP01(2017)103}{{\em JHEP} {\bfseries 01}
  (2017) 103}, \href{http://arxiv.org/abs/1602.05971}{{\ttfamily
  arXiv:1602.05971 [hep-th]}}.

\bibitem{Gomis:2014woa}
J.~Gomis and N.~Ishtiaque, ``{K\"ahler potential and ambiguities in 4d $
  \mathcal{N} $ = 2 SCFTs},''
  \href{http://dx.doi.org/10.1007/JHEP04(2015)169}{{\em JHEP} {\bfseries 04}
  (2015) 169}, \href{http://arxiv.org/abs/1409.5325}{{\ttfamily arXiv:1409.5325
  [hep-th]}}.

\bibitem{Binder:2019jwn}
D.~J. Binder, S.~M. Chester, S.~S. Pufu, and Y.~Wang, ``{$ \mathcal{N} $ = 4
  Super-Yang-Mills correlators at strong coupling from string theory and
  localization},'' \href{http://dx.doi.org/10.1007/JHEP12(2019)119}{{\em JHEP}
  {\bfseries 12} (2019) 119}, \href{http://arxiv.org/abs/1902.06263}{{\ttfamily
  arXiv:1902.06263 [hep-th]}}.

\bibitem{Chester:2020dja}
S.~M. Chester and S.~S. Pufu, ``{Far beyond the planar limit in
  strongly-coupled $ \mathcal{N} $ = 4 SYM},''
  \href{http://dx.doi.org/10.1007/JHEP01(2021)103}{{\em JHEP} {\bfseries 01}
  (2021) 103}, \href{http://arxiv.org/abs/2003.08412}{{\ttfamily
  arXiv:2003.08412 [hep-th]}}.

\bibitem{Chester:2022sqb}
S.~M. Chester, ``{Bootstrapping 4d $\mathcal{N}=2$ gauge theories: the case of
  SQCD},'' \href{http://arxiv.org/abs/2205.12978}{{\ttfamily arXiv:2205.12978
  [hep-th]}}.

\bibitem{Fiol:2020bhf}
B.~Fiol, J.~Mart\'\i{}nez-Montoya, and A.~Rios~Fukelman, ``{The planar limit of
  $\mathcal{N}=2$ superconformal field theories},''
  \href{http://dx.doi.org/10.1007/JHEP05(2020)136}{{\em JHEP} {\bfseries 05}
  (2020) 136}, \href{http://arxiv.org/abs/2003.02879}{{\ttfamily
  arXiv:2003.02879 [hep-th]}}.

\bibitem{Fiol:2020ojn}
B.~Fiol, J.~Martfnez-Montoya, and A.~Rios~Fukelman, ``{The planar limit of $
  \mathcal{N} $ = 2 superconformal quiver theories},''
  \href{http://dx.doi.org/10.1007/JHEP08(2020)161}{{\em JHEP} {\bfseries 08}
  (2020) 161}, \href{http://arxiv.org/abs/2006.06379}{{\ttfamily
  arXiv:2006.06379 [hep-th]}}.

\bibitem{Fiol:2021icm}
B.~Fiol and A.~R. Fukelman, ``{The planar limit of $ \mathcal{N} $ = 2 chiral
  correlators},'' \href{http://dx.doi.org/10.1007/JHEP08(2021)032}{{\em JHEP}
  {\bfseries 08} (2021) 032}, \href{http://arxiv.org/abs/2106.04553}{{\ttfamily
  arXiv:2106.04553 [hep-th]}}.

\bibitem{Fiol:2022vvv}
B.~Fiol and A.~Rios~Fukelman, ``{A derivation of the planar limit of $
  \mathcal{N} $ = 2 chiral correlators},''
  \href{http://dx.doi.org/10.1007/JHEP11(2022)034}{{\em JHEP} {\bfseries 11}
  (2022) 034}, \href{http://arxiv.org/abs/2209.12019}{{\ttfamily
  arXiv:2209.12019 [hep-th]}}.

\bibitem{Dolan:2001tt}
F.~A. Dolan and H.~Osborn, ``{Superconformal symmetry, correlation functions
  and the operator product expansion},''
  \href{http://dx.doi.org/10.1016/S0550-3213(02)00096-2}{{\em Nucl. Phys. B}
  {\bfseries 629} (2002) 3--73},
  \href{http://arxiv.org/abs/hep-th/0112251}{{\ttfamily arXiv:hep-th/0112251}}.

\bibitem{Chester:2020vyz}
S.~M. Chester, M.~B. Green, S.~S. Pufu, Y.~Wang, and C.~Wen, ``{New modular
  invariants in $ \mathcal{N} $ = 4 Super-Yang-Mills theory},''
  \href{http://dx.doi.org/10.1007/JHEP04(2021)212}{{\em JHEP} {\bfseries 04}
  (2021) 212}, \href{http://arxiv.org/abs/2008.02713}{{\ttfamily
  arXiv:2008.02713 [hep-th]}}.

\bibitem{Binder:2018yvd}
D.~J. Binder, S.~M. Chester, and S.~S. Pufu, ``{Absence of $D^4 R^4$ in
  M-Theory From ABJM},'' \href{http://dx.doi.org/10.1007/JHEP04(2020)052}{{\em
  JHEP} {\bfseries 04} (2020) 052},
  \href{http://arxiv.org/abs/1808.10554}{{\ttfamily arXiv:1808.10554
  [hep-th]}}.

\bibitem{Billo:2017glv}
M.~Billo, F.~Fucito, A.~Lerda, J.~F. Morales, Y.~S. Stanev, and C.~Wen,
  ``{Two-point correlators in $N =2$ gauge theories},''
  \href{http://dx.doi.org/10.1016/j.nuclphysb.2017.11.003}{{\em Nucl. Phys. B}
  {\bfseries 926} (2018) 427--466},
  \href{http://arxiv.org/abs/1705.02909}{{\ttfamily arXiv:1705.02909
  [hep-th]}}.

\bibitem{Billo:2018oog}
M.~Billo, F.~Galvagno, P.~Gregori, and A.~Lerda, ``{Correlators between Wilson
  loop and chiral operators in $ \mathcal{N}=2 $ conformal gauge theories},''
  \href{http://dx.doi.org/10.1007/JHEP03(2018)193}{{\em JHEP} {\bfseries 03}
  (2018) 193}, \href{http://arxiv.org/abs/1802.09813}{{\ttfamily
  arXiv:1802.09813 [hep-th]}}.

\bibitem{Billo:2019fbi}
M.~Bill\`o, F.~Galvagno, and A.~Lerda, ``{BPS wilson loops in generic conformal
  $ \mathcal{N} $ = 2 SU(N) SYM theories},''
  \href{http://dx.doi.org/10.1007/JHEP08(2019)108}{{\em JHEP} {\bfseries 08}
  (2019) 108}, \href{http://arxiv.org/abs/1906.07085}{{\ttfamily
  arXiv:1906.07085 [hep-th]}}.

\bibitem{Fiol:2018yuc}
B.~Fiol, J.~Mart\'\i{}nez-Montoya, and A.~Rios~Fukelman, ``{Wilson loops in
  terms of color invariants},''
  \href{http://dx.doi.org/10.1007/JHEP05(2019)202}{{\em JHEP} {\bfseries 05}
  (2019) 202}, \href{http://arxiv.org/abs/1812.06890}{{\ttfamily
  arXiv:1812.06890 [hep-th]}}.

\bibitem{tutte}
W.~T. Tutte, ``{A census of slicings},'' {\em Can. J. Math} {\bfseries 14}
  (1962) 708--722.

\bibitem{Gopakumar:2012ny}
R.~Gopakumar and R.~Pius, ``{Correlators in the Simplest Gauge-String
  Duality},'' \href{http://dx.doi.org/10.1007/JHEP03(2013)175}{{\em JHEP}
  {\bfseries 03} (2013) 175}, \href{http://arxiv.org/abs/1212.1236}{{\ttfamily
  arXiv:1212.1236 [hep-th]}}.

\bibitem{Fiol:2021jsc}
B.~Fiol and A.~R. Fukelman, ``{On the planar free energy of matrix models},''
  \href{http://dx.doi.org/10.1007/JHEP02(2022)078}{{\em JHEP} {\bfseries 02}
  (2022) 078}, \href{http://arxiv.org/abs/2111.14783}{{\ttfamily
  arXiv:2111.14783 [hep-th]}}.

\bibitem{Dorigoni:2021guq}
D.~Dorigoni, M.~B. Green, and C.~Wen, ``{Exact properties of an integrated
  correlator in $ \mathcal{N} $ = 4 SU(N) SYM},''
  \href{http://dx.doi.org/10.1007/JHEP05(2021)089}{{\em JHEP} {\bfseries 05}
  (2021) 089}, \href{http://arxiv.org/abs/2102.09537}{{\ttfamily
  arXiv:2102.09537 [hep-th]}}.

\bibitem{Russo:2013kea}
J.~G. Russo and K.~Zarembo, ``{Massive N=2 Gauge Theories at Large N},''
  \href{http://dx.doi.org/10.1007/JHEP11(2013)130}{{\em JHEP} {\bfseries 11}
  (2013) 130}, \href{http://arxiv.org/abs/1309.1004}{{\ttfamily arXiv:1309.1004
  [hep-th]}}.

\end{thebibliography}\endgroup
\end{document}